\documentclass[referee,usenatbib]{mn2e}

\usepackage{amssymb,amsmath,aas_macros}
\usepackage{graphicx}
\usepackage{multirow}
\usepackage{natbib}

\def\cite{\citep}

\usepackage{bm}
\usepackage{lscape}
\newcommand{\be}{\begin{equation}}
\newcommand{\ee}{\end{equation}}
\def\bea{\begin{eqnarray}}
\def\ena{\end{eqnarray}}

\usepackage{url}
\usepackage{float}
\usepackage{bm}
\footnotesize
\newdimen\minuswidth    %define @ width of minus sign for tables
\setbox0=\hbox{$-$}
\minuswidth=\wd0
\catcode`@=\active
\def@{\kern\minuswidth}
\newdimen\digitwidth    %define ! a one digit width for tables
\setbox0=\hbox{\rm0}
\digitwidth=\wd0
\catcode`!=\active
\def!{\kern\digitwidth}
\normalsize

\usepackage[usenames,dvipsnames]{color}

\newcommand{\email}[1]{\thanks{Email: #1}}

\footnotesize
\newdimen\digitwidth    %define ! a one digit width for tables
\setbox0=\hbox{\rm0}
\digitwidth=\wd0
\catcode`!=\active
\def!{\kern\digitwidth}
\normalsize
\begin{document}
\title[Comparison of pulsar positions from timing and VLBI]
{Comparison of Pulsar Positions from Timing and Very Long Baseline Astrometry} \makeatletter
\author[J.B. Wang et al.]{J. B. Wang,$^{1,2,3}$
	\email{wangjingbo@xao.ac.cn}
	W.A. Coles,$^4$
	G. Hobbs,$^3$
	R.M. Shannon,$^{3,5}$
	R.N. Manchester,$^3$
	\newauthor
	M. Kerr,$^{3,6}$
	J.P. Yuan,$^{1,2}$
	N. Wang,$^{1,2}$
	M. Bailes,$^7$
	N. D. R. Bhat,$^5$
	S. Dai,$^3$
	\newauthor
	J. Dempsey,$^8$
	M. J. Keith,$^9$
	P. D. Lasky,$^{10}$
	Y. Levin,$^{10}$
	S. Os{\l}owski,$^7$
	V. Ravi,$^{11}$
	\newauthor
	D. J. Reardon,$^{10,3}$
	P. A. Rosado,$^7$
	C.J. Russell,$^{12}$
	R. Spiewak,$^7$
	W. van~Straten,$^7$
	\newauthor
	L. Toomey,$^3$
	L. Wen,$^{13}$
	X.-P. You,$^{14}$
	X.-J. Zhu,$^{13}$
	\\
	$^1$Xinjiang Astronomical Observatory, Chinese Academy of Science,
	150 Science 1-Street, Urumqi, Xinjiang, China, 830011 \\
	$^2$Key Laboratory of Radio Astronomy, Chinese Academy of Science, 150 Science 1-Street, Urumqi, Xinjiang, China, 830011 \\
	$^3$CSIRO Astronomy \& Space Science, PO Box 76, Epping, NSW 1710, Australia \\
	$^4$Electrical and Computer Engineering, University of California at San Diego, La Jolla, California, U.S.A. \\
	$^5$International Centre for Radio Astronomy Research, Curtin University, Bentley, Western Australia 6102, Australia\\
	$^6$Naval Research Laboratory, 4555 Overlook Ave., SW, Washington, DC 20375, USA\\
	$^7$Centre for Astrophysics and Supercomputing, Swinburne University of Technology, P.O. Box 218, Hawthorn, Victoria 3122, Australia\\
	$^8$CSIRO Information Management \& Technology, Box 225, Dickson ACT 2602 \\
	$^9$Jodrell Bank Centre for Astrophysics, University of Manchester, M13 9PL, UK.\\
	$^{10}$Monash Centre for Astrophysics, School of Physics and Astronomy, Monash University, VIC 3800, Australia \\
	$^{11}$Cahill Center for Astronomy and Astrophysics, MC 249-17, California Institute of Technology, Pasadena, CA 91125, USA\\
	$^{12}$CSIRO Scientific Computing Services, Australian Technology Park, Locked Bag 9013, Alexandria, NSW 1435, Australia \\
	$^{13}$School of Physics, University of Western Australia, Crawley, WA 6009, Australia\\
	$^{14}$School of Physical Science and Technology, Southwest University, Chongqing, 400715, China\\
}
\date{printed \today}
\maketitle
\begin{abstract}
	\indent Pulsar positions can be measured with high precision using both pulsar timing methods and very-long-baseline interferometry (VLBI). Pulsar timing positions are referenced to a solar-system ephemeris, whereas VLBI positions are referenced to distant quasars.  Here we compare pulsar positions from published VLBI measurements with those obtained from pulsar timing data from the Nanshan and Parkes radio telescopes in order to relate the two reference frames. We find that the timing positions differ significantly from the VLBI positions (and also differ between different ephemerides). A statistically significant change in the obliquity of the ecliptic of $2.16\pm0.33$\,mas is found for the JPL ephemeris DE405, but no significant rotation is found in subsequent JPL ephemerides. The accuracy with which we can relate the two frames is limited by the current uncertainties in the VLBI reference source positions and in matching the pulsars to their reference source.    Not only do the timing positions depend  on the ephemeris used in computing them, but also different segments of the timing data  lead to varying position estimates. These variations are mostly common to all ephemerides, but slight changes are seen at the 10$\mu$as level between ephemerides.

\end{abstract}


\section{Introduction}

Since the early 1980s (e.g., Bartel et al., 1985 and references therein) VLBI observations of pulsars have been providing pulsar positions 
with sub-milliarcsecond (mas) accuracy.    The pulsar timing method is based on observations of pulse times-of-arrival (ToAs) from one or more pulsars. These ToAs are predicted using a model for the astrometric, rotational and orbital properties of each pulsar.  If the position of the pulsar is not included with sufficient accuracy in the timing model then the timing residuals (the differences between the predicted and   ToAs) will include a sine term with a period of a year.  Pulsar timing packages, such as \textsc{tempo2} (Hobbs, Edwards \& Manchester 2006), can be used to fit for the amplitude and phase of the sinusoidal signal and, hence, improve the modelled position of the pulsar.  As ToAs for pulsars with millisecond periods can often be measured with a precision of $\sim$100\,ns the position of such pulsars can also be determined with sub-mas accuracy using these timing methods (see, e.g., Edwards, Hobbs \& Manchester 2006).

Pulsar timing arrays use millisecond pulsars to search for the signatures of ultra-low-frequency gravitational waves (e.g., Shannon et al. 2015) and provide exceptionally precise source positions. These measurements (from the Parkes Pulsar Timing Array project; Manchester et al. 2013) form the basis of the work that we are describing here. Unfortunately there are not many VLBI measurements published for millisecond pulsars because such pulsars are relatively weak.

Younger pulsars with periods of order 1\,s are more common and often brighter than the millisecond pulsars, but it is usually not possible to measure their pulse ToAs with such accuracy and the long-term stability of these pulsars is poorer (see, e.g., Hobbs et al. 2010 for a description of the timing noise in 366 young pulsars). However, there are many VLBI determinations of their positions.   
One of the goals of this project was to determine if inclusion of a much larger number of young pulsars would improve the results that could be obtained using many fewer millisecond pulsars. As we will show, results from the young pulsars were consistent with those from the millisecond pulsars, but they did not improve the frame-tie precision.

Fomalont et al. (1984) compared positions obtained using interferometry and the timing method.  They noted that the differences in the pulsar positions obtained via these two methods could result from many causes including calibration errors in the interferometry and instabilities in the pulsar rotation or errors in the planetary ephemeris for the timing method.  Bartel et al. (1996) computed rotation matrices to transform between various ephemerides, including the JPL planetary ephemeris DE200\footnote{The JPL ephemerides are described in ftp://ssd.jpl.nasa.gov/pub/eph/planets/ioms/ExplSupplChap8.pdf.}.  Since the year 2003 (starting with the DE405 ephemeris) the reference frame for the ephemerides has been the International Celestial Reference Frame (ICRF) as VLBI measurements of planets and spacecraft were used in their construction. 
For completeness we note the following major changes in the ephemerides since DE405\footnote{based on \url{ftp://ssd.jpl.nasa.gov/pub/eph/planets/README.txt}}:
\begin{itemize}
	\item DE414: Created in 2005 this included an upgraded orbit of Pluto and ranging data from the MGS and Odyssey spacecraft. 
	\item DE421: Created in 2008 this ephemeris included new fits to planetary and lunar laser ranging data
	\item DE430: From 2013, this was referenced to the ICRF 2.
	\item DE432: This was created in 2014 specifically for the New Horizons mission for Pluto.  It did not include effects relating to nutation.
	\item DE435: Created in 2016 for the Cassini project and primarily provides an updated orbit for Saturn\footnote{see \url{ftp://ssd.jpl.nasa.gov/pub/eph/planets/ioms/de435.iom.pdf} for details.}.
\end{itemize}
Pulsar timing observations analyzed with these ephemerides therefore should agree with VLBI observations. The most recent analysis was carried out by Madison et al. (2013) who did not detect a significant rotation between the frames.  Here we extend their work and we find statistically significant differences in the obliquity of the ecliptic for the more recent JPL solar system ephemerides DE405 through 435 compared to ICRF2, but show that these differences are dominated by the VLBI and timing position of a single pulsar, PSR~J0437$-$4715.  When we update the position of this pulsar with more recent information,  we only find a statistically significant difference for DE405.

As described by Madison et al. (2013), this type of work is of special importance for pulsar timing.  The use of VLBI measurements in pulsar timing arrays would improve the sensitivity of the data sets both for gravitational wave searches
and for searches for unknown solar-system bodies (such as the postulated planet X; see e.g., Batygin \& Brown 2016).

In order to ensure that the work presented here is reproducible we have made our data files and MATLAB code publically available. Details are provided in Appendix B.

\section{Observations}
The timing data sets were obtained from the Nanshan 25m-diameter telescope (see Wang et al. 2001 for an introduction to the Nanshan observing system) and the Parkes 64m-diameter telescope (see Reardon et al. 2016 for a description of the Parkes data sets that we used here).  All the VLBI pulsar positions are from published work (Bartel et al. 1996; Brisken et al. 2002, 2003; Chatterjee et al. 2001, 2004, 2009; Deller et at. 2008, 2009, 2012, 2013, 2016).
%In this work the uncertainties of the observations and the parameters estimated from them are denoted as $\sigma$ which is the estimated standard deviation of the observation or parameter, so the 68\% confidence interval is $\pm\sigma$.
Throughout this paper, uncertainties are denoted as $\sigma$ and represent 68\% confidence intervals.

\subsection{Timing data analysis}

%Timing data sets were obtained from the Nanshan 25m-diameter telescope (see Wang et al. 2001 for an introduction to the Nanshan observing system) and the Parkes 64m-diameter telescope (see Reardon et al. 2016 for a description of the Parkes data sets that we used here).  

Regular timing observations of 74 pulsars using the Nanshan telescope of the Xinjiang Astronomical Observatory commenced in 2000 January in the 18\,cm observing band. A cryogenic receiver system has been used since July 2002. The total bandwidth of 320\,MHz is divided into 128 channels for each circular polarization and digitized at 1\,ms intervals.  The observation time for each pulsar is usually between 4 and 16 minutes. We selected from these pulsars the 25 for which VLBI positions have been published.

The Parkes pulsar timing array project (PPTA) has, since 2005, been regularly timing around 20 millisecond pulsars in the 10, 20 and 50\,cm observing bands (Manchester et al. 2013; Reardon et al. 2016). We selected the five for which VLBI positions have been published.
We have used the PPTA observations after they have been combined with earlier timing observations in the 20\,cm band for some of the pulsars (Verbiest et al. 2009). Our timing analysis is based on the \textsc{tempo2} software package.  As described by Hobbs, Edwards \& Manchester (2006) this package corrects all known phenomena at the $\sim$1\,ns level.  This requires accounting for variations in the Earth's rotation.  These are accounted for within \textsc{tempo2} using the C04 series of the Earth Orientation Parameters (EOP) that are published by the International Earth Rotation Service (IERS). Details are available in McCarthy \& Petit (2004).  \textsc{tempo2} also requires the telescope positions on the Earth (which are determined with high accuracy in geocentric coordinates) and many other effects. Details are provided in the Hobbs, Edwards \& Manchester (2006) and Edwards, Hobbs \& Manchester (2006) series of papers about the \textsc{tempo2} software.

The earlier observations cannot be corrected for fluctuations in the interstellar medium (ISM) because they are at a single band but they can be analyzed together with the new ISM-corrected observations using the ``split-Cholesky" method discussed by Reardon et al. (2016). We followed their method and used their noise models to re-run 
%This data set has been recently analyzed by Reardon et al. (2016) and so we did not have to change the noise models. However, we did re-run 
the timing analysis using different planetary ephemerides. We also extracted the covariance matrix of each pulsar's right ascension (RA) and declination (Dec), which was not published by Reardon et al. (2016) and changed the timing position epoch to match that of the VLBI observation. It should be noted that the proper motion of many pulsars greatly exceeds the uncertainty in position over the data span, so the measurement of the proper motion is important. When the VLBI measurement is at the edge of the timing span over even outside it, the uncertainty in the proper motion can be the primary limitation.

The Nanshan pulsars were reanalyzed and new noise models obtained. The white noise was estimated using the {\sc efacEquad} plugin in the {\sc tempo2} package to rescale the ToA uncertainties so that they better represent the observed scatter in the residuals (Wang et al. 2015). The red noise was estimated using the {\sc spectralModel} plugin (Coles et al. 2011). The epoch of the timing position for each pulsar was set to be the same as that of the VLBI measurement.  As before, we obtained the astrometric parameters using a range of different planetary ephemerides and we obtained the covariance matrix of RA and Dec for each analysis.

\subsection{VLBI data}
%All the VLBI pulsar positions are from published work (Bartel et al. 1996; Brisken et al. 2002, 2003; Chatterjee et al. 2001, 2004, 2009; Deller et at. 2008, 2009, 2012, 2013, 2016). 
The most recent available observations of PSR~J1939+2134 (Bartel et al. 1996) also provide a weighted average of their position with previous observations. We use this weighted average as it reduces the uncertainties by about 50\% and improves the overall analysis by about 10\%.
In most cases the authors provide the formal RA and Dec uncertainties including the known positional uncertainties in the calibration sources used. However in the case of PSR J0437$-$4715 the uncertainties were given independently. In this case we summed the quoted differential uncertainty and the uncertainty in the calibrator source in quadrature. However it was completely dominated by the uncertainty in the calibrator position. In fact all the VLBI position uncertainties are dominated by the ICRF reference sources. In most cases the VLBI positions were referenced to ICRF1 (Ma et al. 1998). The observations of PSR~J1939+2134 were referenced to the International Earth Rotation Service 1993 reference frame. Since this (IERS, 1993) was the immediate precursor to ICRF1 we have treated them as the same. The same calibration sources are now available in ICRF2 (Fey et al. 2009) so we have corrected the source positions to ICRF2. We did not correct the published uncertainties to those in ICRF2 except for PSR~J0437$-$4715 because we cannot separate the calibration error from other errors. 
%Our VLBI position uncertainties may therefore be underestimated by as much as 50\% for pulsars other than PSR J0437$-$4715. 
It is clear that the VLBI position uncertainties are likely to be underestimated, but by exactly how much is difficult to quantify and will vary between pulsars.  When converting from ICRF1 to ICRF2 we typically halve the error in the position of the reference source, but the remaining errors remain unchanged.  For our work we assume that the position uncertainties may be underestimated by as much as 50\% for pulsars other than PSR~J0437$-$4715, but we do not attempt to further quantify such errors.
We note that the root-mean-square (rms) difference between ICRF1 and ICRF2 is $\sim$0.2\, mas for our millisecond pulsars. The rotation between ICRF1 and ICRF2 has been estimated by Fey (2009) who found it consistent with zero with an uncertainty of $\sim$0.02\,mas. 

\subsection{The effects of ephemerides and ICRF frames}

\begin{figure}
	\centerline{\includegraphics[width=4in]{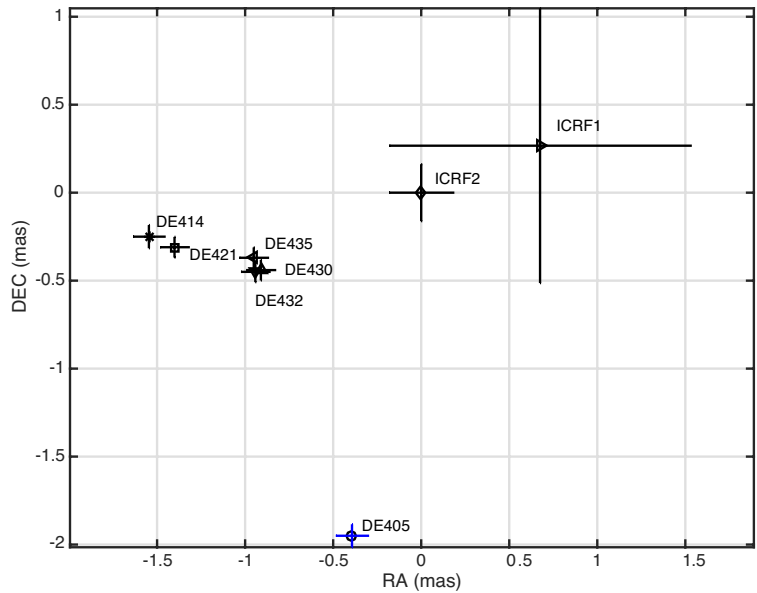}}
	\caption{Positions from VLBI and timing for J0437$-$4715 with respect to the VLBI position under ICRF2. Error bars on VLBI are 1$\sigma$, but those on timing are 10$\sigma$ for visibility.
	}\label{fg:0437}
\end{figure}

Timing positions depend on the planetary ephemeris used. The strongest millisecond pulsar, PSR~J0437$-$4715, shows this clearly. In Figure~\ref{fg:0437}, 
we plot the apparent position according to VLBI referenced to both ICRF1 and ICRF2.
We also plot the apparent positions of this pulsar from pulsar timing using planetary ephemerides DE405 through DE435 (Standish 1998; Folkner et al. 2008). 
DE405 was the first ephemeris to be referenced to ICRF1 and DE430 was the first referenced to ICRF2.
The error bars on the timing positions are extremely small and have been increased in the figure by a factor of 10 to be visible. 
All timing positions determined with different ephemerides differ by much more than their statistical uncertainties, but positions determined using DE405 are particularly  inconsistent. This pattern also applies to the other millisecond pulsars in our data set.
In general the timing position uncertainty is dominated by ephemeris differences just as the ICRF uncertainty dominates the VLBI position uncertainty.

Since ICRF2 positions are available for all VLBI calibrators we use them as the reference throughout the rest of this paper. These VLBI positions for the five millisecond pulsars we used are listed in Table~\ref{tb:vlbi}. The differential positions (with respect to ICRF2) are given in Table~\ref{tb:jpl} for VLBI under ICRF1 and for timing with ephemerides DE405 through 435. The timing measurements of RA and Dec for PSR~J1022+1001 are highly correlated and the uncertainties reflect this. It was necessary to carry six significant figures in the cross correlation to analyze this pulsar. We note that the ecliptic longitude is well-defined and this pulsar does contribute significantly to the frame tie.

The young pulsars are listed in Table~\ref{tb:young}. Even though we calculated positions with respect to all the planetary ephemerides, for brevity we only tabulate our results with respect to JPL DE430, the most recent major adjustment of the JPL ephemerides.  Position determinations for these young pulsars are not sufficiently precise to show differences between ephemerides.

\begin{table*}
	\caption{Published VLBI positions for millisecond pulsars corrected to ICRF2. The epoch of the observation is given in modified Julian days (MJDs). The values in parentheses are the 1 $\sigma$ errors in the least significant digit.}\label{tb:vlbi}
	\begin{tabular}{lllllll}
		\hline
		PSR       &          Right Ascension    &        $\sigma_{\rm RA}$  &   Declination    &       $\sigma_{\rm Dec}$  & Epoch & Ref \\
	    & & (mas) & & (mas) & (MJD)  &\\
		\hline
		J0437$-$4715 & 04:37:15.883205 &  0.18  &  $-$47:15:09.03213 &  0.16  &  54100.0 & Deller et al. (2008) \\
		J1022$+$1001&  10:22:57.995715 &  1.5 &  +10:01:52.76475 &  1  & 56000.0 & Deller et al. (2016) \\
		J1713$+$0747 & 17:13:49.530594 &  1.5 &  +07:47:37.51915 &  2 &  52275.0 & Chatterjee et al. (2009) \\
		J1939$+$2134 & 19:39:38.561182 &  2.7 &  +21:34:59.13155 &  2.4  & 46853.0 & Bartel et al. (1996) \\
		J2145$-$0750 & 21:45:50.458788 &  1.5 &  $-$07:50:18.51295 &  2  & 56000.0 & Deller et al. (2016) \\
		\hline
	\end{tabular}
\end{table*}

\begin{table*}

	\caption{Published VLBI and new timing positions for millisecond pulsars. Positions are differential with respect to the VLBI ICRF2 position in Table 1. Epochs are as in Table 1.  All positions and uncertainties are in mas. The dimensionless normalized cross correlation between RA and Dec is given by $C_{\rm cor}$.
	}\label{tb:jpl}
	\begin{tabular}{llrlrlr}
		\hline
		PSR   &     Type   &   RA    &   $\sigma_{\rm RA}$   &   Dec    &    $\sigma_{\rm Dec}$   & $C_{\rm cor}$  \\
		\hline
		J0437$-$4715 &  ICRF1 &   0.680 &   0.86 &   0.270  &  0.78 & -\\
	       &   DE405  &  $-$0.393 &   0.0092 &  $-$1.950  &   0.0065  &   0.017\\
	&  DE414 &  $-$1.547 &   0.0090 &  $-$0.250  &   0.0066 &  $-$0.032\\
 & DE421 &  $-$1.401  &   0.0082 &  $-$0.310  &   0.0059 &   0.005\\
 & DE430 &  $-$0.911 &   0.0082 &  $-$0.440  &   0.0059 &   0.005\\
 &  DE432 &  $-$0.942  &   0.0082 &  $-$0.450  &   0.0059 &   0.005\\
 &  DE435 &  $-$0.951  &   0.0082&  $-$0.370  &   0.0059 &   0.005\\ \\
		J1022+1001 &  ICRF1 &  $-$0.224  &   1.500  &   0.250 &   1.000 & - \\
	     &   DE405 &  22.415 &  19.383  &  57.580  &  49.065  &   0.999957\\
      &   DE414 &  15.126  &  16.191  &  41.180  &  40.998  &   0.999973\\
      &   DE421 &  16.809  &  19.374  &  45.990  &  49.044  &   0.999957\\
      &   DE430 &  15.333  &  16.227  &  40.940  &  41.087  &   0.999973\\
      &   DE432 &  15.453  &  16.223 &  41.230  &  41.078  &   0.999973\\
      &   DE435 &  23.292  &  18.696  &  61.910  &  47.329  &   0.999965\\ \\
		J1713+0747 &  ICRF1 &   0.085 &   1.500  &  $-$0.153  &   2.000  &-\\
	     &   DE405 &   2.840 &   0.031  &   4.120  &   0.065  &   0.230\\
      &   DE414 &   1.616 &   0.031 &   2.450  &   0.065  &   0.230\\
      &   DE421 &   1.719  &   0.035  &   2.490  &   0.073  &   0.234 \\
      &   DE430 &   2.063 &   0.035  &   2.620  &   0.074 &   0.230 \\
      &   DE432 &   2.033 &   0.035  &   2.630 &   0.074 &   0.230 \\
      &   DE435 &   1.988 &   0.035  &   2.540  &   0.074 &   0.230 \\ \\
		J1939+2134 &  ICRF1 &  -0.276  &   2.700  &  -0.046  &   2.400 &-\\ 
	     &   DE405 &  $-$0.095 &   0.341 &   4.470  &   0.455 &  $-$0.031\\
      &   DE414 &  $-$1.689  &   0.341  &   3.300  &   0.455  &  $-$0.031\\
      &   DE421 &  $-$1.509  &   0.341  &   3.310  &   0.455  &  $-$0.031\\
      &   DE430 &  $-$1.236  &   0.341  &   3.270  &   0.455  &  $-$0.031\\
      &   DE432 &  $-$1.295 &   0.341  &   3.280  &   0.455  &  $-$0.031\\
      &   DE435 &  $-$1.307 &   0.341  &   3.190  &   0.455  &  $-$0.031\\ \\
		%          &   DE405 &  $-$1.865 &   0.341 &   6.020  &   0.455 &  $-$0.031\\
		%          &   DE414 &  $-$3.459  &   0.341  &   4.850  &   0.455  &  $-$0.031\\
		%          &   DE421 &  $-$3.279  &   0.341  &   4.860  &   0.455  &  $-$0.031\\
		%          &   DE430 &  $-$3.006  &   0.341  &   4.820  &   0.455  &  $-$0.031\\
		%          &   DE432 &  $-$3.065 &   0.341  &   4.830  &   0.455  &  $-$0.031\\
		%          &   DE435 &  $-$3.077 &   0.341  &   4.740  &   0.455  &  $-$0.031\\
		% J1939+2134 -0.094500  2.721401  4.470000  2.442695 DE405
		%J1939+2134 -1.689000  2.721461  3.300000  2.442817 DE414
		%J1939+2134 -1.509000  2.721459  3.310000  2.442813 DE421
		%J1939+2134 -1.236000  2.721459  3.270000  2.442813 DE430
		%J1939+2134 -1.294500  2.721459  3.280000  2.442813 DE432
		%J1939+2134 -1.306500  2.721461  3.190000  2.442813 DE435
		J2145$-$0750 &  ICRF1 &   0.179  &   1.500  &  -0.053  &   2.000 &- \\
	       &   DE405 &   1.796 &   0.578  &   0.860  &   1.555 &  $-$0.957\\
	&   DE414 &   0.720  &   0.578  &   0.530  &   1.555 &  $-$0.957\\
 &   DE421 &   0.651  &   0.660 &   0.640  &   1.773 &  $-$0.957\\
 &   DE430 &   1.004 &   0.660 &   0.540  &   1.773 &  $-$0.957\\
 &   DE432 &   1.001 &   0.660 &   0.560  &   1.772  &  $-$0.957\\
 &   DE435 &   0.917 &   0.660 &   0.480  &   1.773 &  $-$0.957\\
		\hline
	\end{tabular}
\end{table*}

\begin{table*}

	\caption{Published VLBI ICRF2 and new timing positions for young pulsars. Timing positions are with respect to 
	VLBI positions and are given in mas.}\label{tb:young}
	\begin{tabular}{rrrrrrrl}
		\hline
		Pulsar & P0 &\multicolumn{2}{c}{VLBI} &Epoch&\multicolumn{2}{c}{Timing - VLBI (mas)}\\
		       &      (s) &  RA$_{\mbox{ICRF2}}$     & Dec$_{\mbox{ICRF2}}$     & (MJD)     & RA$_{\mbox{DE430}}$     & Dec$_{\mbox{DE430}}$      &Ref\\
		\hline
		J0034$-$0721& 0.9429&00:34:08.8703(1)&$-$07:21:53.409(2)& 52275.0  &$-$446(73500) & $-$4609(154000)   & 1 \\ %Chatterjee et al. (2009)\\
		J0108$-$1431&           0.8075&01:08:08.347016(88)&$-$14:31:50.187139(1069) &54100.0            &$-$195(360) & 723(640)   & 2 \\ %Deller et al.(2009) \\
		J0139+5814& 0.2724&01:39:19.7401(12)&+58:14:31.819(17)& 52275.0   &$-$223(181)&14(120)&1 \\ %Chatterjee et al. (2009)\\
		J0332+5434&0.7145&03:32:59.3862(10)&+54:34:43.5051(150)& 51544.0   &$-$136(30)&103(27)   & 3\\%Brisken et al. (2002)\\
		J0358+5413&0.1563&03:58:53.71650(330) & +54:13:13.7273(50)&51544.0    &2.7(50)&12(11) &1 \\ \\ %Chatterjee et al. (2009)\\
		J0452$-$1759&0.5489&04:52:34.1057(1)&$-$17:59:23.371(2)&52275.0  &49(153) & $-$31(221)   &1 \\ %Chatterjee et al. (2009)\\
		J0454+5543&0.3407&04:54:07.7506(1)  &+55:43:41.437(2) & 52275.0 &$-$33(32)&23(32) &1 \\ %Chatterjee et al. (2009)\\
		J0538+2817&             0.1431& 05:38:25.0572(1) &+28:17:09.161(2) & 52275.0  &3(225)& $-$1539(2100)   &1 \\ %Chatterjee et al. (2009)\\
		J0630$-$2834&1.2444& 06:30:49.404393(43) &$-$28:34:42.778813(372)&54100.0 &$-$33(50)&$-$33(64)&2 \\ %Deller et al. (2009) \\
		J0659+1414&0.3849& 06:59:48.1472(7)& +14:14:21.160(10)&51544.0&32(315)&$-$840(1700)& 4\\ \\% Brisken et al. (2003)\\
		J0814+7429&1.2922&08:14:59.5412(10) &+74:29:05.3671(150)& 51544.0                &-243(285)&-170(86)   & 2\\%Brisken et al. (2002) \\
		J0820$-$1350& 1.2381& 08:20:26.3817(1) &$-$13:50:55.859(2) &52275.0  &26(53)&$-$56(84)   &1 \\ %Chatterjee et al. (2009)\\
		J0953+0755&0.2530&09:53:09.3071(10)  &+07:55:36.1475(150) & 51544.0 &$-$25(63)&$-$83(161)   & 3\\%Brisken et al.(2002)\\
		J1136+1551&1.1879&11:36:03.1829(10) &+15:51:09.7257(150)& 51544.0    &50(21)& 91(34)   & 3\\%Brisken et al.(2002)\\
		J1239+2453&1.3824&12:39:40.3589(10)  &+24:53:50.0194(150)  &  51544.0 &12(28)& 49(37)   & 3\\ \\%Brisken et al.(2002)\\ 
		J1509+5531&0.7396&15:09:25.6298(1) &+55:31:32.394(2) & 52275.0   &$-$66(69)&7(38) &1 \\ %Chatterjee et al. (2009)\\
		J1543+0929&0.7484&15:43:38.8250(1) &+09:29:16.339(2)  & 52275.0   &$-$45 (180) & 209(320)&1 \\ %Chatterjee et al. (2009)\\
		J1932+1059&0.2265&19:32:13.94970(30)  &+10:59:32.4198(50)&51544.0   &78(17) & $-$34(29)   & 5\\%Chatterjee et al. (2004)\\
		J1935+1616&0.3587& 19:35:47.8259(1) & +16:16:39.986(2) & 52275.0    &$-$6.6(127) & $-$17(19)   & 5\\%Chatterjee et al. (2004)\\
		J2018+2839&0.5579& 20:18:03.8332(10) &+28:39:54.1564(150) & 51544.0   &11(60) & 137(68)   & 3\\ \\%Brisken et al.(2002)\\
		J2022+2854&0.3434& 20:22:37.0712(10) &+28:54:23.0337(150)  & 51544.0  &$-$14(26) & 1(28)   & 3\\%Brisken et al.(2002)\\
		J2022+5154&0.5291&20:22:49.8655(10)  &+51:54:50.3811(150) & 51544.0   &$-$3(63) &$-$23(43)   & 3\\%Brisken et al.(2002)\\
		J2048$-$1616&1.9615&20:48:35.640637(40)  & $-$16:16:44.553501(147) &  52275.0     &460(405)&346(1400) &2 \\ %Deller et al. (2009)\\
		J2157+4017&1.5252&21:57:01.8495(1)  & +40:17:45.986(2)  & 52275.0    &$-$111(92)& $-$36(80)   &1 \\ %Chatterjee et al. (2009)\\
		J2313+4253&0.3494& 23:13:08.6209(1) & +42:53:13.043(2) & 52275.0   &$-$7.5(825)& 12.6(78) &1 \\ %Chatterjee et al. (2009)\\
		\hline
		\multicolumn{8}{l}{(1) Chatterjee et al. (2009); (2) Deller et al. (2009); (3) Brisken et al (2002); (4) Brisken et al (2003); (5) Chatterjee et al (2004)}\\
	\end{tabular}
\end{table*}

\section{Estimation of the Frame-Tie Rotation}

To find the rotation that relates the celestial and ecliptic frames we follow the procedure described by Madison et al. (2013), keeping the same notation (their section 5.1). 
Since the rotation angles and the measurement errors are small ($< 10^{-8}$ rad) the transformation can be linearized with no sacrifice in accuracy. We assume that the observations, both timing and VLBI, are provided in equatorial coordinates, $\bm{\theta}_{\rm eq}  = ( \alpha ,  \delta)$ (where $\alpha$ and $\delta$ are the right ascension and declination respectively), and we convert these into direction cosines $\hat{\bm{n}}_{\rm eq} = (x_{\rm eq}, y_{\rm eq}, z_{\rm eq})$ 
\begin{equation}
	x_{\rm eq}=\cos{\alpha} \cos{\delta}; \;\;
	y_{\rm eq}=\sin{\alpha} \cos{\delta}; \;\;
	z_{\rm eq}=\sin{\delta}.
\end{equation}
The differentials can be written as $d\hat{\bm n}_{\rm eq} = \mathbf{D_{\rm eq} }\ d{\bm \theta}_{\rm eq}$ where
\begin{equation}
	\mathbf{D}_{\rm eq} =
	\left( \begin{array}{cc}
			-\sin{\alpha}\cos{\delta} & -\cos{\alpha}\sin{\delta}  \\
			\cos{\alpha}\cos{\delta}& -\sin{\alpha}\sin{\delta}   \\
			0 & \cos{\delta}
	\end{array} \right).
\end{equation}
The position vectors in the ICRF (IC) and timing (DE) frames (both column vectors) can be related by a rotation matrix, $\mathbf{\Omega}$, as shown below.
\begin{equation}
	 {\hat{\bm n}_{\rm DE}} = \mathbf{\Omega_{IC}^{DE}}  {\hat{\bm n}_{\rm IC}}
	%\mathbfit{\hat{n}_{\rm DE}} = \mathbfss{\Omega_{IC}^{DE}} \mathbfit{\hat{n}_{\rm IC}}
\end{equation}
For the very small angles considered here the rotation matrix can be written as $\mathbf{\Omega^{DE}_{IC} = I + \Lambda}$ where
$\mathbf{I}$ is the identity matrix and
\begin{equation}
	\Lambda =
\left( \begin{array}{ccc}
0 & A_z & -A_y \\
 -A_z& 0 & A_x \\
 A_y & -A_x & 0
\end{array} \right).
\end{equation}
Here $A_x$, $A_y$ and $A_z$ are the (right-handed) Euler rotation angles about the x, y, and z axes respectively. Note that the x axis is towards the vernal equinox
%intersection of the ecliptic and equatorial planes is the x axis, 
so $A_x$ represents an offset in the obliquity of the ecliptic.

In matrix form we can rewrite the offset between the timing and VLBI positions as $d \hat{\bm n}_{\rm DE} - d \hat{\bm n}_{\rm IC}  = \mathbf{B A} + {\bm \epsilon}_n$. Here 
\begin{equation}
	\mathbf{B} =
\left( \begin{array}{ccc}
0 & -z & y \\
 z & 0 & -x \\
 -y & x & 0
\end{array} \right),
\end{equation}
The direction cosines ($x$,$y$,$z$) in the equation above could be either the timing or VLBI coordinates as they agree to $\sim$9 significant figures.
$\mathbf{A}$ is the column vector $(A_x, A_y, A_z)$ and ${\bm \epsilon}_n$ is the vector of the uncertainties in the difference between the two position vectors. The covariance matrix of $d \hat{\bm n}$ is
$\mathbf{D_{\rm eq}} \rm{Cov} ({\bm \theta}_{\rm eq}) \ \mathbf{D_{\rm eq}^T}$ and $\Sigma$, the covariance matrix of ${\bm \epsilon}_n$, is the sum of
the covariance matrix of $d \hat{\bm n}_{\rm DE}$ and that of $d \hat{\bm n}_{\rm IC}$.

This is a linear least squares problem and is easily solved if  $\Sigma$
is known. Madison et al. (2013) made the simplifying assumption that $\Sigma$ is
diagonal and computed the variances on the diagonal by propagating the errors from ${\bm \theta}_{\rm eq}$ to
$\hat{\bm n}_{\rm eq}$ using our Equation (1).  This assumption creates two problems. $\Sigma$ is not diagonal and would not be diagonal even if ${\rm Cov}({\bm \theta}_{\rm eq})$ were diagonal. Furthermore $\Sigma$ is singular and cannot be inverted as required by the solution given by Equation (12) in Madison et al. (2013). 

These problems can be minimized by extracting the full ${\rm Cov}({\bm \theta}_{\rm eq})$ from the {\sc{tempo2}} fit, computing $\Sigma$ from ${\rm Cov}({\bm \theta}_{\rm eq})$ and using the Moore-Penrose pseudo-inverse in place of $\Sigma^{-1}$ (see equation 18 in Zhu, et al., 2014). Unfortunately we do not have the full ${\rm Cov}({\bm \theta}_{\rm eq})$ for the VLBI measurements, but these are never as highly correlated as are the timing positions. 

We tested this procedure, but we also found a more straightforward method in which we altered the least squares equations
$d \hat{\bm n}_{\rm DE} - d \hat{\bm n}_{\rm IC}  = \mathbf{B A} + {\bm \epsilon}_n$ to avoid the singularity in the covariance matrix $\Sigma$.
This was done by making the substitution
$\mathbf{D_{eq}} \ ( d{\bm \theta}_{\rm DE} - d{\bm \theta}_{\rm IC}) = \mathbf{B A }+{\bm \epsilon}_n$, then pre-multiplication of both sides by $\mathbf{D_{eq}^T}$. New least squares equations are found by premultiplication of both sides by $\mathbf{(D_{eq}^T D_{eq})^{-1}}$.
The new least squares problem applies directly to the observations rather than the transformed observations:
\begin{equation}
 \ ( d{\bm \theta}_{\rm DE} - d{\bm \theta}_{\rm IC}) =  \mathbf{(D_{eq}^T D_{eq})^{-1} D_{eq}^T B A } + \epsilon_{\theta}.
\end{equation}
Here $\mathbf{D_{eq}^T D_{eq}}$ is diagonal and is trivially inverted. The product $\mathbf{(D_{eq}^T D_{eq})^{-1} D_{eq}^T}$ is the pseudo inverse of $\mathbf{D_{eq}}$. The results from this method are numerically identical to the original method.

One of the pulsars in our set, PSR~J1022+1001, is very close to the ecliptic, and so $\alpha$ and $\delta$ are highly correlated. This pulsar would be better analyzed in ecliptic coordinates because the ecliptic latitude and longitude are almost uncorrelated. However we found that we could solve the least squares problem in equatorial coordinates using double precision. We did this to avoid potential confusion in defining the ecliptic reference frame.

\section{Results for the frame-tie rotation}

Prior to applying our method to our data, we tested our procedure with the positions of the two millisecond pulsars
(PSRs J0437$-$4715 and J1713+0747) that were processed by Madison et al. (2013). We found, as expected, negligible difference between our rotation matrix and theirs. In this case, with only two pulsars, there only four data points and there are three parameters to be obtained. Since there is only one degree of freedom, the problem is more like an inversion than a least squares fit and the weighting of the observations has very little effect on the resulting estimate.

Below we first present our results using only the five millisecond pulsars and then demonstrate the effect of including the young pulsars.

\subsection{Millisecond Pulsars}

In Table~\ref{tb:angles}, we present the results for all five millisecond pulsars, using six different JPL ephemerides. There is no significant rotation about the y or z axes in any ephemeris, but there is a consistently significant rotation about the x axis, $A_x$, i.e., an offset in the obliquity of the ecliptic. This is $2.4 \pm 0.3$\,mas in DE405, but was significantly corrected in DE415 and has remained at $\sim 0.8 \pm$ 0.3\,mas in more recent ephemerides. 
The normalized $\chi^2$ is consistently slightly below unity, probably because the VLBI positions were slightly improved by re-referencing them to ICRF2, 
but, as noted earlier, the published errors were not reduced when making this correction. Overall the $\chi_{norm}^2$ indicates that $\rm{Cov} ({\bm \theta}_{\rm eq})$ describes the uncertainty in the observations 
${\bm \theta}_{\rm eq}$ reasonably well.

\begin{table*}
	\caption{Rotation angles (in mas) between VLBI positions and timing positions using different JPL ephemerides for five millisecond pulsars.}\label{tb:angles}
\begin{tabular}{llllllll}
\hline
Ephermeris & $A_x$  &  $\sigma_{A_x}$ &  $A_y$  &  $\sigma_{A_y}$  & $A_z$ &   $\sigma_{A_z}$  & $\chi_{norm}^2$   \\
\hline
DE405 & 2.36 & 0.27  & 0.67 & 0.61  & $-$1.19 & 0.70 & 0.93  \\
DE414 & 0.76  & 0.26 & 1.26 & 0.58  & $-$0.02 & 0.66 & 0.84  \\
DE421 & 0.77 & 0.26  & 1.11  & 0.59  & $-$0.02 & 0.67 & 0.86 \\
DE430 & 0.87 & 0.26 & 0.99  & 0.58 & $-$0.43 & 0.66  & 0.83  \\
DE432 & 0.88  & 0.26 & 1.01 & 0.58 & $-$0.42 & 0.66 & 0.83 \\
DE435 & 0.76  & 0.28 & 0.92  & 0.63 & $-$0.27 & 0.71 & 0.98  \\
\hline
\end{tabular}
\end{table*}

\begin{figure}
\centerline{\includegraphics[width=4.5in]{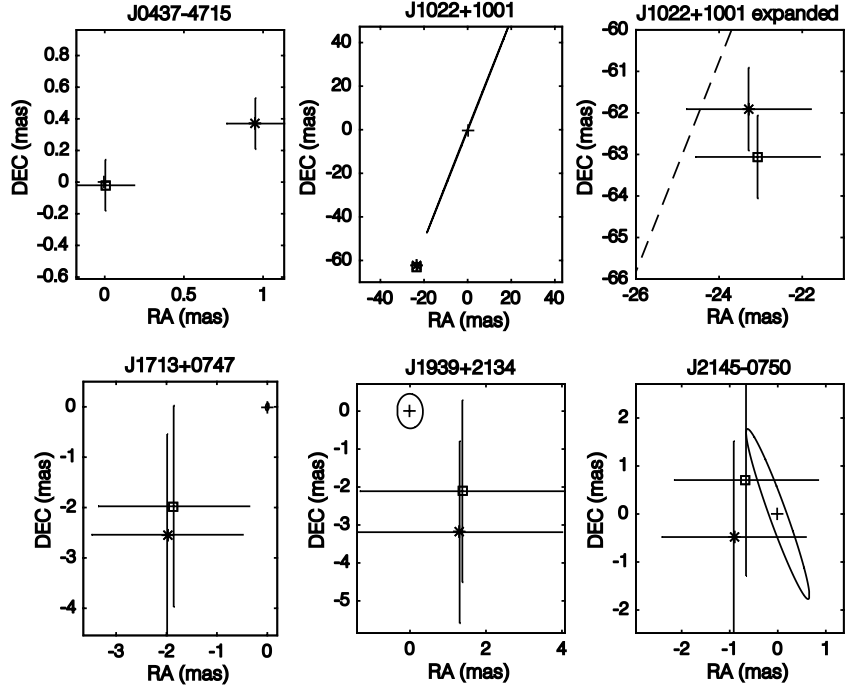}}
\caption{Pulsar positions from timing (+), original VLBI(asterisk) and transformed VLBI(square) for millisecond pulsars. 
After transformation the `square' should match the `+'. The timing errors are shown as an ellipse to illustrate the effect of the correlation between
RA and Dec. The timing uncertainty for J0437-4715 and J1713+0747 is too small to be visible on the plot. The timing uncertainty ellipse for J1022+1001
has almost degenerated to a line. It is extended 30\% in the expanded panel to show that the VLBI positions are within 1$\sigma$ of the timing position.}\label{fg:msp}
\end{figure}

The positions before and after rotation are shown in Figure~\ref{fg:msp} for the most recent ephemeris, DE435. The timing position marked with an `+' is defined to be the origin. The original VLBI position in ICRF2 is marked with an `x', and after rotation it is marked with a square. One can see that the fitting process puts the transformed VLBI position of PSR~J0437$-$4715 very close to its timing position.  The VLBI position of the other pulsars are also transformed to be consistent, within an error bar, with their timing positions, but the error bars are much larger. PSR~J0437$-$4715 is the brightest millisecond pulsar in the sky so both its timing and VLBI positions are exceptionally well-measured and it dominates the fit. 

To further understand the dominance of PSR~J0437$-$4715 we removed it from the array and repeated the fit. This increased both  $A_x$ and $\sigma_{Ax}$, to 2.8 and 1.5\,mas respectively.  However when we remove the other pulsars from the fit, one at a time, we find that each contributes significantly to reducing the uncertainty, $\sigma_{Ax}$. Indeed the rotation $A_x$ did not exceed $3\sigma_{Ax}$ comfortably unless all five pulsars were included. Because the result depends so strongly on PSR~J0437$-$4715 we discussed this VLBI measurement with the lead author of the corresponding publication (Deller, private communication, 2017). He noted that continuing unpublished VLBI observations suggest that there is a calibration error of $\sim +0.8$\,mas in the published observation of PSR~J0437$-$4715. Accordingly we recomputed the frame tie assuming this bias and increasing the uncertainty by 0.8\,mas in quadrature because the bias is not yet well understood. The revised results, presented in Table~5, no longer show a significant rotation (except for the DE405 ephemeris).
We will use this correction to the VLBI position of PSR~J0437$-$4715 in the remainder of this work.

\begin{table*}
	\caption{Rotation angles (in mas) after correction
	for a suspected bias in the right ascension of PSR~J0437$-$4715.}\label{tb:revangles}
\begin{tabular}{llllllll}
\hline
Ephermeris & $A_x$  &  $\sigma_{A_x}$ &  $A_y$  &  $\sigma_{A_y}$  & $A_z$ &   $\sigma_{A_z}$  & $\chi_{norm}^2$   \\
\hline
DE405 & 2.16 & 0.33 & 0.13 & 0.78 & -1.36 & 0.71 & 0.92 \\
DE414 & 0.57 & 0.31 & 0.73 & 0.74 & -0.19 & 0.67 & 0.83 \\ 
DE421 & 0.58 & 0.32 & 0.59 & 0.76 & -0.18 & 0.68 & 0.85  \\
DE430 & 0.67 & 0.31 & 0.47 & 0.74 & -0.59 & 0.67 & 0.82  \\
DE432 & 0.69 & 0.31 & 0.49 & 0.74 & -0.58 & 0.67 & 0.82  \\
DE435 & 0.57 & 0.34 & 0.40 & 0.81 & -0.44 & 0.73 & 0.97  \\
\hline
\end{tabular}
\end{table*}
%Obviously the frame-tie cannot be determined by PSR~J0437$-$4715 alone, nor can it be determined from the other four pulsars together. Knowing which pulsars contribute is important as the obliquity of the ecliptic is thought to have been determined with an uncertainty of 0.25 mas (W. Folkner, private communication, 2017) using VLBI measurements of Mars and the spacecraft Mars Odyssey and Mars Reconnaissance Orbiter (Folkner and Border, 2015).   As noted in Deller et al. (2008), the published uncertainty for the VLBI position of PSR~J0437$-$4715 does not attempt to include the systematic effect of phase referencing over a $\sim$2 degree angular separation, meaning the VLBI uncertainty quoted here (despite being dominated by the uncertainty of the reference source J0439$-$4922, which should be the largest source of uncertainty) is likely to still be under-estimated.  Future VLBI observations designed specifically to constrain accurately the absolute position could both better estimate this uncertainty and reduce its magnitude substantially (A. Deller, priv. comm.)  Even though we believe that the mathematical formalism presented here is correct, we note that our result is only significant at the 3$\sigma$ level and depends strongly on the VLBI position of a single pulsar.
%One can see that PSRs~J0437$-$4715 and J1713$+$0747 contribute the most, but the other pulsars are also useful contributors.

\subsection{Inclusion of Young Pulsars}

We were able to include all the young pulsars although their positional uncertainties are as much as $10^5$ times larger than those of the millisecond pulsars. We did not expect all to contribute, but including the very noisy pulsars provided a test of the numerical stability of the algorithm. 
We did find a small problem in estimating the uncertainty. As is the normal practice we scale the covariance matrix of the estimated parameters by the normalised $\chi^2$-value ($\chi_{norm}^2$). However adding young pulsars sometimes reduced the $\chi_{norm}^2$ below unity and caused the covariance matrix to be underestimated. We modified the scaling so that it is only applied when the $\chi_{norm}^2$ is greater than unity which solved this problem. The result of including all the young pulsars is given in Table~\ref{tb:de430}. One can see that 
%including the very noisy young pulsars degrades the sensitivity a few percent, but that 
they are completely consistent with the millisecond pulsars. 
%The uncertainty in the young pulsars is probably slightly overestimated (i.e. a bit conservative). 

\begin{table*}
	\caption{Rotation angles (in mas) under DE430 and ICRF2 for five millisecond pulsars and all pulsars in the VLBI sample}\label{tb:de430}
\begin{tabular}{llllllll}
\hline
Data & $A_x$  &  $\sigma_{A_x}$ &  $A_y$  &  $\sigma_{A_y}$  & $A_z$ &   $\sigma_{A_z}$  & $\chi_{norm}^2$  \\
\hline
%5  & 0.8866 & 0.2772 & 1.0422 & 0.6273 & $-$0.4871 & 0.7105 & 0.9300 \\
%All pulsars & 0.8997  & 0.2859 &  1.0739 &  0.6455 &  $-$0.5220 &  0.7310 &  0.8700 \\
5 MSPs & 0.67 & 0.31 & 0.47 & 0.74 & -0.59 & 0.67 & 0.82  \\
All pulsars & 0.68  & 0.28 &  0.49 &  0.63 &  $-$0.63 &  0.71 &  0.86 \\
%DE430 & 0.67 & 0.31 & 0.47 & 0.74 & -0.59 & 0.67 & 0.82  \\
%0.6777 0.2808 0.4852 0.6325 -0.6305 0.7149 0.8584 all

\hline
\end{tabular}
\end{table*}

We also searched for outliers among the young pulsars by appending them, one at a time, to the five millisecond pulsars. The resulting estimates of $A_x$ had a mean and standard deviation of $0.68 \pm 0.01$\,mas. As the standard deviation of $A_x$ is 0.3 mas the variation of $A_x$ when including different young pulsars is completely negligible. We conclude that young pulsars would not bias a frame tie, but they will not be useful 
%in estimating a frame tie 
unless the timing error can be greatly improved. Simply including more young pulsars will not help. 

\subsection{Effect of Assuming that the Covariance Matrix is Diagonal}

We use the positions determined with the JPL DE405 ephemeris to provide a basis for comparing different assumptions about the covariance matrix of the observations.  We chose this ephemeris  are there is a detectable frame-tie rotation in $A_x$ that provides a ``test signal".
We compare the results obtained using three different assumptions: (1) our algorithm, (2) our algorithm neglecting the cross correlation between RA and Dec and (3) the Madison et al. (2013) algorithm which assumes that the matrix $\Sigma$ is diagonal. We then exclude the two pulsars near the ecliptic (PSRs J1022+1001 and J2145$-$0750) and repeat the comparison. The results are in Table~\ref{tb:alg}.

When we exclude the near-ecliptic pulsars cases (1) and (2) are identical because the RA and Dec are almost uncorrelated. In this comparison case (3) is similar to cases (1) and (2) but the $\chi_{norm}^2$ and all the other errors are underestimated because there are only 3 degrees of freedom in the residuals and the algorithm assumes six degrees of freedom. When we include the two near-ecliptic pulsars the results are more complex. Case (2) gets the signal $A_z$ reasonably well but overestimates $A_y$ and $A_z$. In fact $A_z$ becomes marginally significant. Case (3) is worse, as one might expect. It gets the signal reasonably well but underestimates the error, with $A_z$ becoming a 3$\sigma$ detection.

In summary, the need for including the off-diagonal terms increases when they become larger. In this particular data set it became important for two of the five millisecond pulsars. This is easily done for the timing measurements. We urge VLBI observers to make an effort to specify the correlation between their right ascension and declination measurements.

%Comparing our full algorithm with five and with three pulsars, we see that the rotation $A_x$ remains significant although both the $A_x$ and $\sigma_{A_x}$ become considerably larger with fewer pulsars. In addition an apparently significant rotation appears in $A_y$ with fewer pulsars. This shows the importance of using the ecliptic pulsars and the value of adding still more pulsars when the VLBI measurements become available.
%
%Comparing our full algorithm with algorithms 2 and 3, we see that failure to include the cross correlation of RA and Dec causes a significant error when ecliptic pulsars are present, but is otherwise negligible. When ecliptic pulsars are not present the Madison et al. (2013) algorithm more than doubles the uncertainty, but when ecliptic pulsars are present it breaks down completely.

\begin{table*}
	\caption{Comparison of the rotation angles (in mas) between DE405 and ICRF2 with three different algorithms.}\label{tb:alg}
\begin{tabular}{lrrrrrrl}
\hline
\multicolumn{8}{c}{Using all five millsecond pulsars}\\
Alg. & $A_x$  &  $\sigma_{A_x}$ &  $A_y$  &  $\sigma_{A_y}$  & $A_z$ &   $\sigma_{A_z}$  & $\chi_{norm}^2$   \\
\hline
%1 & 0.87 & 0.26 & 0.99  & 0.58 & $-$0.43 & 0.66  & 0.83  \\
%2 & 1.26  & 0.31  & 2.06 & 0.76  & $-$1.65  & 0.87 & 0.69  \\
%3 & $-$13.48 & 17.12 & $-$38.37 & 7.87  & $-$12.55  & 4.81  & 3536 \\
%EPHEM  Ax    Axerr   Ay    Ayerr   Az    Azerr   MSE 
%DE405 2.1558 0.2703 0.1296 0.6101 -1.3568 0.6898 0.9156 		our algorithm
%DE405 2.4728 0.3127 0.9878 0.7600 -2.3433 0.8651 0.7717 		assuming C diagonal
%DE405 2.2045 0.1905 0.2881 0.4507 -1.5362 0.5056 0.6566 		Madison
1 & 2.1558 &  0.2703 &  0.1296 &  0.6101 &  $-$1.3568 &  0.6898 &  0.9156 \\	%	our algorithm
2 & 2.4728 &  0.3127 &  0.9878 &  0.7600 &  $-$2.3433 &  0.8651 &  0.7717 \\	%	assuming C diagonal
3 & 2.2045 &  0.1905 &  0.2881 &  0.4507 &  $-$1.5362 &  0.5056 &  0.6566 \\	%	Madison

\hline
\multicolumn{8}{c}{Excluding two millisecond pulsars near the ecliptic plane}\\
\hline
%1 & 1.63  & 0.42 & 3.04  & 1.07  & $-$2.78  & 1.23 & 0.62 \\
%2 & 1.63  & 0.42 & 3.04  & 1.07  & $-$2.78  & 1.23 & 0.62 \\
%3 & 2.98  & 0.93  & 6.02  & 2.29 & 0.89 & 1.06 & 6.33  \\
%EPHEM  Ax    Axerr   Ay    Ayerr   Az    Azerr   MSE 
%DE405 2.6657 0.4304 1.5024 1.0990 -2.9308 1.2575 0.7794 	our algorithm
%DE405 2.6658 0.4305 1.5027 1.0993 -2.9312 1.2578 0.7797 	assuming C diagonal
%DE405 2.5191 0.2716 1.1276 0.6964 -2.4937 0.7939 0.4832  Madison 
1 & 2.6657 &  0.4304 &  1.5024 &  1.0990 &  $-$2.9308 &  1.2575 &  0.7794 \\ % 	our algorithm
2 & 2.6658 &  0.4305 &  1.5027 &  1.0993 &  $-$2.9312 &  1.2578 &  0.7797  \\ %	assuming C diagonal
3 & 2.5191 &  0.2716 &  1.1276 &  0.6964 &  $-$2.4937 &  0.7939 &  0.4832 \\ %  Madison 

\hline
\multicolumn{8}{c}{Alg1=this work; Alg2=RA \& Dec uncor.; Alg3=Madison et al. (2013)}
\end{tabular}
\end{table*}

\subsection{Effect of the Number of Pulsars}

\begin{figure}
\centerline{\includegraphics[width=3.5in]{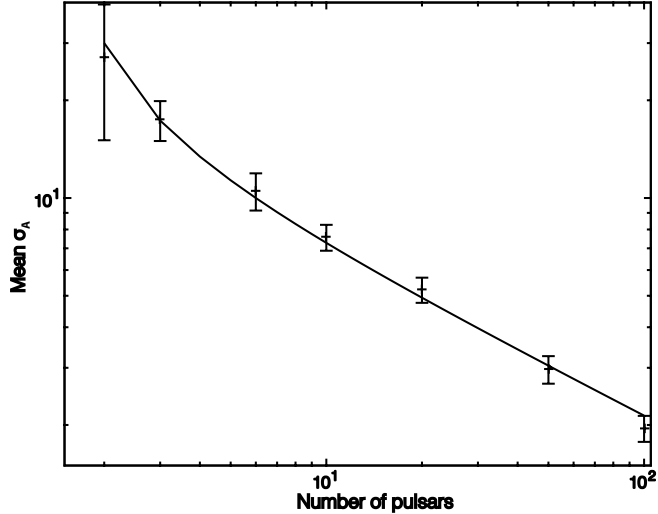}}
\caption{$\bar{\sigma}_A$ (mas) versus the number of pulsars derived from a simulation.}\label{fg:simu}
\end{figure}

As adding more young pulsars will not improve the frame tie,  here we estimate the improvement that could be gained by adding more millisecond pulsars.  
If we had $N_p$ pulsars with equal,  uncorrelated errors in RA and Dec, then we would have $2 N_p$ independent measurements from which we need to estimate the three independent parameters (the three Euler angles). Accordingly we would expect the variance of each Euler angle to decrease as $1/(2N_p - 3)$. We tested this by placing $N_p =$\,2, 5, 10, 20, 50  and 100 pulsars at uniformly distributed random positions on the sky. As we have assumed that the noise in RA and Dec is independent we explicitly ignore the possibility that some pulsars may be near the ecliptic. We assumed a 10-mas rms independent Gaussian error in the differential
RA and Dec.  We then calculated the rotation angles using our method. We simulated 20 realizations of the sky for each $N_p$, and 1000 realizations
of the noise for each realization of the sky. For each sky realization we found the rms rotation angles over the 1000 simulations, $\sigma_{A_x}$ etc.
From these we determined the rms rotation angle for each sky realization
$\sigma_A = (\sigma_{A_x}^2 + \sigma_{A_y}^2 + \sigma_{A_z}^2)^{0.5}$. 
The mean of $\sigma_A$ over the 20 sky realizations is plotted as a function of $N_p$  in Figure~\ref{fg:simu}. The error bars on Figure~\ref{fg:simu} are the rms of $\sigma_A$ over the sky realizations.
The line drawn through the simulated measurements is $\sigma_A = 30 {\rm mas}/(2 N_p-3)^{0.5}$ and matches well with our simulations.

We can use this equation to estimate how many pulsars would be required to halve the uncertainty in our estimate of the obliquity. 
%Two of our five millisecond pulsars are close to the ecliptic so they really only contribute one independent measurement between them. 
Two of our five millisecond pulsars are close to the ecliptic so they together effectively only contribute one independent measurement.
Accordingly we currently have four ``typical" millisecond pulsars.  We would need roughly 11 such millisecond pulsars to halve our current rms uncertainty.

\section{Time variations in the timing positions}

Our ability to perform the frame tie is limited by the precision of the VLBI positions, but the timing positions are far more precise and clearly show the effects of different ephemerides. This suggests that a search for time variations in the apparent timing positions might provide a useful measure of the accuracy of the ephemerides because the timing positions (at a reference epoch) should not vary with time. 
We have done this for the two bright pulsars, PSRs~J0437$-$4715 and J1713+0747.
Obtaining accurate timing positions requires several years of observation so we break our data into  blocks of four years overlapping by two years. 
These data have been corrected for fluctuations in ISM dispersion where possible, but the early data could not be corrected because there was only data available in the 20 cm observing band.

The mean positions for both pulsars vary significantly with the ephemeris used (DE405, 414, 421, and 430) but the time variation in PSR~J1713+0747 is not statistically significant and that in PSR~J0437$-$4715 is only marginally significant. The time variations are highly correlated between the ephemerides, so we have taken an average of the most recent (DE414, 421 and 430) as a reference.  This reference position is plotted in the top panel of Figure~4 for PSR~J0437$-$4715 and similarly in Figure~5 for PSR~J1713+0747.
We have estimated how much position variation could occur in the early observations of PSR~J0437$-$4715 because of uncorrected fluctuations in the ISM and it is about a factor of 10 too weak to explain the observed variation (our calculation is presented in Appendix A). On the other hand the points between 2000 and 2004 which exceed 2$\sigma$ are not statistically independent because the blocks overlap by half.  We conclude that the observed time variation is probably simply measurement error.

\begin{figure}
\centerline{\includegraphics[width=7in]{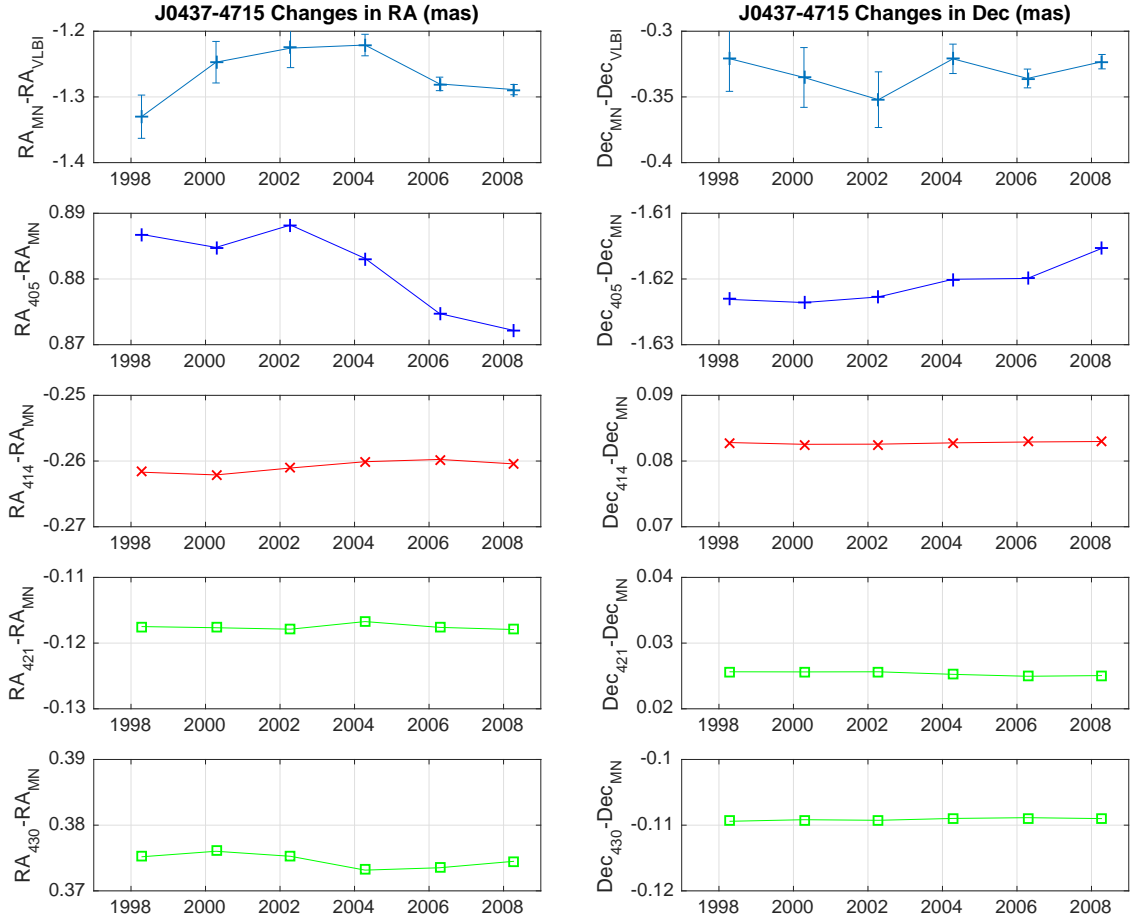}}
\caption{Temporal changes in timing position for PSR J0437$-$4715. The top panels gives the average of positions determined using 
ephemerides DE414, 421 and 430.
The lower panels give the positions using DE405, 414, 421 and 430 with respect to the average in the top panels. }%\label{fg:rat}
\end{figure}
However the measurement error is the same for all ephemerides, so we can search for smaller differences between the ephemerides.
We compare the positions determined by each ephemeris with the reference position and plot these in the lower 4 panels of Figures 4 and 5. 
These show much smaller variation, of order 10 $\mu$as. As the variations between ephemerides are highly correlated the errors are also highly correlated and the differences are much more significant that would be suggested by the error bars on the top panel. 
The variation in DE405 in both pulsars appears to be real and this would be reasonable because we can easily detect frame tie errors in this ephemeris. It is not clear that the variations in the more recent ephemerides are real although the fluctuations in DE421 appear to be smallest in both pulsars. 

%As the observations are the same and only the ephemeris changes, the estimates are highly correlated. This allows very small variations to be seen clearly. One can see that DE421 shows the least position variation.
%
%Observations of the position of PSR J1713+0747, in the same format as Figure 4, are shown in Figure 5. Here one can see that the fluctuations in the reference position are similar in magnitude to those of PSR~J0437$-$4715, probably because they are both low dispersion pulsars. The differential fluctuations with respect to the reference are also similar in magnitude to those of PSR~J0437$-$4715 and again DE421 shows the least fluctuation.

\begin{figure}
\centerline{\includegraphics[width=7in]{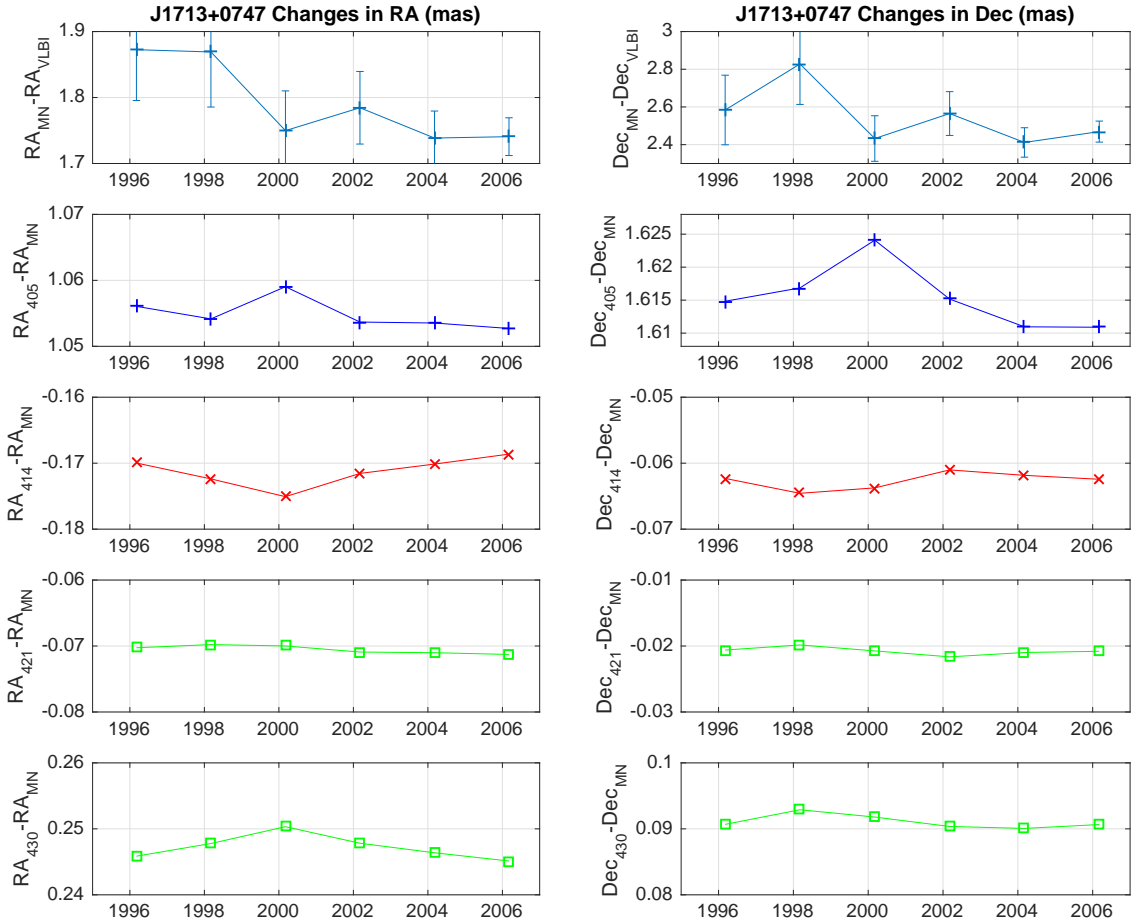}}
\caption{Temporal changes in timing position for PSR J1713+0747 in the same format as Figure 4. }%\label{fg:rat2}
\end{figure}

%In the top panel the position using ephemerides DE405 is shown. The lower panels show the differences between positions measured with DE414, 421 and 430 and those in the top panel. It is apparent that there are significant temporal variations in all ephemerides but that there are very small temporal differences between them. Using the Kolmogorov statistical model for the ISM density fluctuations used by Reardon et al (2016) we estimate that the rms uncorrected position variation would be of the order of 1 mas on a time scale of a few years. The early observations are not corrected for fluctuations in dispersion but the recent observations are corrected. Thus the variations seen in the top panels of Figures 4 and 5 are very likely due to the ISM.
%
%The mean differences between ephemerides are large ($\sim 1$ mas) but the time variation in the differences is small ($\sim10 \mu as$ ). Note that the error bars on the differential measurements are computed assuming that the two measurements differenced are statistically independent, but this is not correct. They are highly correlated, but we don't have a way of estimating the correlation coefficient. We believe that the apparent variation of $\sim10\mu as$  in the differential measurements is real. 

The variations between pulsar positions determined with different ephemerides, both in the mean and the temporal variations, demonstrate the sensitivity of pulsar timing and its potential for improving planetary ephemerides. 
%However following this up would take us far from the focus of this paper.

\section{Discussion and conclusions}
 
Data contributing to the JPL ephemerides are dominated by ranging observations between solar-system bodies and have only recently begun to include VLBI measurements of the apparent position of solar-system bodies which can tie the celestial frame of VLBI observations to the ecliptic coordinates of the heliosphere  (Folkner \& Border 2015). The present accuracy of the frame tie is believed to be $\sim0.25$ mas (Folkner, private communication 2017).
So it is not surprising that we find an error in the obliquity of the ecliptic of 2.16$\pm0.33$ mas in DE405, but that any errors in subsequent ephemerides are not statistically significant. However our work shows that pulsar timing observations can be used to detect differences between ephemerides of order $10\mu$as which suggests that they may be useful in refining future ephemerides.

We would like to use the VLBI position, proper motion and parallax in the pulsar timing models in order to remove the requirement to fit for these astrometric parameters.  Currently such fitting removes all signals with a frequency near 1\,yr$^{-1}$ and much of the power near 2\,yr$^{-1}$. Such signals potentially include ephemeris errors,  variations in the relativistic correction from terrestrial to barycentric time and gravitational waves from individual supermassive binary black holes. However our work shows that
present VLBI positions are not yet accurate enough to improve the frame tie or to improve the sensitivity of pulsar timing arrays.

In the near future improvements in the frame tie can be expected primarily from VLBI observations because there are already
$\sim$50 precisely-timed millisecond pulsars in the International Pulsar Timing Array (IPTA, Verbiest et al. 2016). It will take many years to add more  millisecond
pulsars to this set. However we expect that significantly more VLBI observations of these pulsars will soon become available, 
e.g. through the PSR$\pi$ project (Deller et al. 2016).
It is also possible that the Large European Array for Pulsars (LEAP) project could provide astrometric information on a some of these millisecond pulsars.  

In the more distant future, more pulsars will be precisely timed with new instruments and, hopefully, some will be bright enough that precise VLBI positions can be determined.
Pulsar timing programs are planned for both the Five Hundred Metre Spherical Telescope (FAST) currently being commissioned in China and the Square Kilometre Array (SKA).

\section{Acknowledgements}

The Parkes radio telescope is part of the Australia Telescope, which is funded by the Commonwealth of Australia for operation as a National Facility managed by the Commonwealth Scientific and Industrial Research Organisation (CSIRO). This work is supported by West Light Foundation of CAS (No.XBBS201322) and National Natural Science Foundation of China (Nos. 11403086, U1431107, 1137300), the Strategic Priority Research Programme (B) of the Chinese Academy of Sciences (No. XDB230102000) and GH acknowledge support from the Australian Research Council (ARC) Future Fellowship.  We acknowledge comments on early drafts of the manuscript from Dr A. Deller. This work used the astronomy \& astrophysics package for Matlab (Ofek 2014).

\appendix
\section{Expected position variations from turbulent phase gradients in the ISM}

A transverse phase gradient in the ISM would cause a change in the apparent position of a pulsar. Such gradients will vary statistically due to the natural spectrum of turbulent variations in electron density. The resulting phase variations $\phi(\bm{r})$ can be described statistically by the phase structure function $D_p (\bm{s}) = \langle (\phi(\bm{r}+\bm{s}) - \phi(\bm{r}))^2 \rangle$ where the angle brackets denote an ensemble average (You et al, 2007, Keith et al, 2013). Accordingly we can estimate the rms phase gradient over a spatial scale s, $\nabla \phi = \sqrt{D_p (\bm{s})} / |{\bm s}|$ and the resulting angular shift is given by $\bm{\theta}_s = \nabla \phi / k$ where $k = 2 \pi F / c$. The ISM models used by Keith et al. (2013) are given as a time delay structure function $D_t (\bm{s}) = D_p (\bm{s}) /(2 \pi F)^2$
where $\bm{s} = \bm{V}T$ and $D_t(\bm{s} = \bm{V} \times 1000~{\rm days})$ is tabulated. Here $\bm{V}$ is the velocity of the line-of-sight with respect to the ISM.

The velocity $\bm{V}$ and the apparent position shift of the pulsar ${\bm \theta}_p$ depend on the location of the scattering region. We define the fractional distance from the pulsar to the scattering region as $\zeta$. Then we have $V = V_p (1-\zeta)$ and $\theta_p = \theta_s \zeta$. Here $V_p$ is the proper motion of the pulsar. Finally we have
\begin{equation}
{\rm RMS}(\theta_p ) = \sqrt{Dt(1000d) } \ c\  \zeta /  [ (1-\zeta) V_p 1000d].
\end{equation}

Putting in values for PSR~J0437$-$4715 of $D_t (1000) = 1.6 \mu s^2$ (Keith et al, 2013), $V_p = 105$ km/s (Reardon et al. 2016), and $\zeta = 0.22$ (Bhat, Coles, private communication, 2015) we obtain $\theta_p = 0.0024$ mas. Thus the probability that this process (with an rms of 0.0024\,mas) produces the observed single oscillation of $\sim$0.1\,mas peak to peak is low. 

\section{Reproducing our work}

Our software scripts and the input data files are available from the CSIRO Data Access Portal (DAP; \url{https://data.csiro.au}).   The input pulsar timing data files for the millisecond pulsars are available from \url{http://doi.org/10.4225/08/561EFD72D0409}.  The data were processed using \textsc{tempo2} using the different solar system ephemerides described in this paper.  The resulting \textsc{tempo2} positions are recorded in a text file \textsc{msp\_timing}. The VLBI positions are obtained from the literature and stored in \textsc{msp\_vlbi}.  The calibration source positions for the millisecond pulsars are stored in \textsc{cal.txt}.  Similar files exist for the analysis of the young pulsars.  We then use \textsc{MATLAB} to obtain the rotation angles.  All these files can be obtained from the CSIRO DAP (http://doi.org/10.4225/08/58a0e1593c5be). Our \textsc{MATLAB} code requires the \textsc{convertdms.m}, \textsc{coco.m}, \textsc{rotm\_coo.m}, \textsc{cosined.m} and \textsc{obliquity.m} files from the \textsc{Astromatlab} library and \textsc{errorbare.m} a plot routine for drawing error bars\footnote{The errorbare.m file can be obtained from \url{https://au.mathworks.com/matlabcentral/fileexchange/23465-enhanced-errorbar-function/content/errorbare/errorbare.m} and the \textsc{Astromatlab} library is described by Ofek (2014) and available from \url{https://webhome.weizmann.ac.il/home/eofek/matlab/}.}.

\end{document}